\documentclass[prd,a4paper]{revtex4}
\usepackage{graphicx}% Include figure files
\usepackage{dcolumn}% Align table columns on decimal point
\usepackage{bm}% bold math
\usepackage{epsfig}
\newcommand{\be}{\begin{equation}}
\newcommand{\ee}{\end{equation}}
\newcommand{\bea}{\begin{eqnarray}}
\newcommand{\eea}{\end{eqnarray}}
\newcommand{\bean}{\begin{eqnarray*}}
\newcommand{\eean}{\end{eqnarray*}}

\newcommand{\gapproxeq}{\lower
.7ex\hbox{$\;\stackrel{\textstyle >}{\sim}\;$}}
\newcommand{\lapproxeq}{\lower
.7ex\hbox{$\;\stackrel{\textstyle <}{\sim}\;$}}

\begin{document}

\bibliographystyle{unsrt}

\title{\bf Isospin violation in $\phi, J/\psi, \psi^\prime \to \omega \pi^0$ via hadronic loops}

\author{Gang Li$^1$, Qiang Zhao$^{1,2}$, and Bing-Song Zou$^{1,3}$}

\affiliation{1) Institute of High Energy Physics, Chinese Academy
of Sciences, Beijing 100049, P.R. China}

\affiliation{2) Department of Physics, University of Surrey,
Guildford, GU2 7XH, United Kingdom}

\affiliation{3) CCAST, Beijing 100080, P.R. China}

\date{\today}

\begin{abstract}

In this work, we study the isospin-violating decay of $\phi\to
\omega\pi^0$ and quantify the electromagnetic (EM) transitions and
intermediate meson exchanges as two major sources of the decay
mechanisms. In the EM decays, the present datum status allows a
good constraint on the EM decay form factor in the vector meson
dominance (VMD) model, and it turns out that the EM transition can
only account for about $1/4\sim 1/3$ of the branching ratio for
$\phi\to \omega\pi^0$. The intermediate meson exchanges,
$K\bar{K}(K^*)$ (intermediate $K\bar{K}$ interaction via $K^*$
exchanges), $K\bar{K^*}(K)$ (intermediate $K\bar{K^*}$
rescattering via kaon exchanges), and $K\bar{K^*}(K^*)$
(intermediate $K\bar{K^*}$ rescattering via $K^*$ exchanges),
which evade the naive Okubo-Zweig-Iizuka (OZI) rule, serve as
another important contribution to the isospin violations. They are
evaluated with effective Lagrangians where explicit constraints
from experiment can be applied. Combining these three
contributions, we obtain results in good agreement with the
experimental data. This approach is also extended to
$J/\psi(\psi^\prime)\to \omega\pi^0$, where we find contributions
from the $K\bar{K}(K^*)$, $K\bar{K^*}(K)$ and $K\bar{K^*}(K^*)$
loops are negligibly small, and the isospin violation is likely to
be dominated by the EM transition.

\end{abstract}

\maketitle


\vspace{1cm}

\section{Introduction}

The isospin breaking decay channel $\phi \to \omega\pi^0$ has been
measured by experiment with improved
precisions~\cite{Achasov:1999jc}, and the Particle Data Group
quote $BR(\phi \to \omega \pi^0) = (5.2_{-1.1}^{+1.3})\times
10^{-5}$ as the world average for its branching
ratio~\cite{pdg2006}. This decay channel is very interesting due
to the presence of the OZI-rule violation and isospin symmetry
breaking together. These two mechanisms, which generally account
for different aspects of the underlying dynamics, are correlated
in this channel. With the available of much improved experimental
information about other related transitions, one can pursue a
quantitative study of the underlying dynamics and learn more about
the correlation between the OZI-rule violation and isospin
symmetry breaking in the non-perturbative regime.

The electromagnetic (EM) decay of $\phi\to \omega\pi^0$ is an
important source of isospin violations, where the $s$ and
$\bar{s}$ annihilate into a virtual photon, which then decays into
$\omega\pi^0$. The other source of isospin violation originates
from the mass differences between the $u$ and $d$
quark~\cite{Miller:1990iz}. It can contribute to $\phi\to
\omega\pi^0$ via OZI-rule-violating strong decays.

In the literature the isospin violation in $\phi\to \omega\pi^0$
was studied by isoscalar and isovector mixing, e.g.
$\phi$-$\omega$-$\rho^0$ and $\eta^\prime$-$\eta$-$\pi^0$
mixings~\cite{bramon-81,donoghue,ametller,coon-86,coon-87,genz}.
This scenario contains both EM and strong transitions in an
$s$-channel, and allow the $\phi\to\omega\pi^0$ decay without
violating the OZI-rule~\cite{Karnakov:1986sn,Achasov:1990at}. In
such an approach, the EM and strong decays cannot be separated
out.  An alternative view is to separate the EM and strong
processes by explicitly introducing the EM amplitude as an
$s$-channel process, and then including the hadronic loop
contributions as the $t$-channel processes. This will be our focus
in this work. Our strategy is to constrain the EM transition
first, and a well-defined EM transition will then allow us to make
a reliable evaluation of the strong isospin violation mechanism.

The EM transitions can be studied in the vector meson dominance
(VMD) model. Recently, a systematic investigation of the role
played by the EM transitions in $J/\psi(\psi^\prime)\to VP$, where
$V$ and $P$ denote light nonet vector and pseudoscalar mesons,
respectively, was reported in Refs.~\cite{Zhao:2006gw,Li:2007ky},
and the up-to-date experimental data provided a good constraint on
the VMD model. For $\phi\to\omega\pi^0$, the VMD approach has
great advantages: on the one hand, the $\phi$ and $\omega$ meson
masses are very close to the $\rho$ mass. Hence, the EM form
factors can be constrained by the precise data for the $\rho^0$
meson mass and width~\cite{pdg2006}. On the other hand, since
other heavier vectors are rather far away from this kinematic
region, their contributions to the form factor will be limited.
The dominant mechanisms can thus be clarified. The availability of
experimental information for $\phi\to \gamma\pi^0$ and
$\rho\pi+\pi^+\pi^-\pi^0$~\cite{Aloisio:2003ur} is also an
advantage for quantifying the EM contributions.

The isospin-violating strong decay can be related to the OZI-rule
violation at low energies via intermediate hadronic loops as
proposed by Lipkin~\cite{Lipkin:1986bi,Lipkin:1986av}. Microscopic
interpretation of such a scenario as a mechanism for the OZI-rule
violation was investigated by Geiger and Isgur in a quark
model~\cite{Geiger:1992va,isgur-thacker-2001}. For instance, an
$s\bar{s}$ pair of $1^-$ can couple to non-strange $n\bar{n}\equiv
(u\bar{u}+d\bar{d})/\sqrt{2}$ via $K\bar{K}$, $K^*\bar{K}+c.c.$,
etc. Suppressions of such an OZI-rule-violating process come from
the cancellations between the intermediate meson loops and
off-shell effects on the intermediate
states~\cite{lipkin-zou,wu-zhao-zou}. Qualitatively, at high
energies, where the mass scale of the intermediate states becomes
unimportant, one would expect a ``perfect cancellation" among all
those intermediate states, and it recovers the OZI rule. At low
energies, where the mass scale of the individual states is
dominant, the ``perfect cancellation" will break down due to e.g.
$m_u\neq m_d$ originated from the chiral symmetry breaking. The
OZI-rule violations hence give rise to the recognition of isospin
symmetry breakings.

Such a mechanism in $\phi\to\omega\pi^0$ decay can be described as
follows: In $\phi\to\omega\pi^0$, the intermediate charged and
neutral kaon loop transitions are supposed to cancel out if the
isospin symmetry is conserved. However, due to small mass
differences between the $u$ and $d$ quarks, the charged and
neutral kaons will also have small differences in mass, i.e.
$m_{K^0}-m_{K^\pm}=3.972\pm 0.027$ MeV~\cite{pdg2006}, and they
are coupled to the $\phi$ meson with slightly different strength.
The hadronic loops will then have ``imperfect" cancellations and
lead to measurable isospin violating branching ratios. This drives
us to investigate the contributions from the intermediate meson
exchanges to $\phi\to\omega\pi^0$, which are not only an OZI-rule
violating mechanism, but also a source of isospin violations.

A reasonable approach is that at hadronic level, we study the EM
and hadronic loop contributions coherently with the aid of the
up-to-date experimental data. It will enable us to quantify these
two isospin violating sources with some obvious advantages: i) At
hadronic level, we can extract couplings from independent
experimental measurements without knowing all the details about
the quark distribution functions. This technique has been broadly
applied to the study of non-perturbative long-range interactions
in the hadronic decays of heavy quarkonia, especially in
charmonium
decays~\cite{Lu:1992xd,Li:1996yn,Zhao:2005ip,Zhao:2006cx,Liu:2006dq}.
ii) Adopting the experimental constraints on the meson masses and
effective couplings, we also avoid the details about how the
difference of the $u$-$d$ quark masses leads to the corrections to
the decay constants.

In the next Section, we first analyze the EM $\phi$ decay in a VMD
model and then present our intermediate-meson-exchange model with
effective Lagrangians. The numerical results for
$\phi\to\omega\pi^0$ are given in Section III. An extension of
this approach to $J/\psi(\psi^\prime)\to\omega\pi^0$ is also
discussed. A summary is then given in Section IV.

\section{The model}

\subsection{Electromagnetic decay in VMD model}

The $V\gamma^*$ coupling is described by the VMD
model~\cite{Bauer:1975bw},
\be
{\cal L}_{V\gamma}=\sum_V \frac{e M_V^2}{f_V} V_\mu A^\mu \ ,
\ee
where $eM_V^2/f_V$ is a direct photon-vector-meson coupling in
Feynman diagram language, and the isospin 1 and 0 component of the
EM field are both included. It should be noted that this form of
interaction is only an approximation and can have large off-shell
effects arising from either off-shell vector meson or virtual
photon fields. In this approach we consider such effects in the
$V\gamma P$ coupling form factor which will then be absorbed into
the energy-dependent widths of the vector mesons.

The typical effective Lagrangian for the $V\gamma P$ coupling is:
\be\label{lagrangian-1} {\cal L}_{V\gamma P}=\frac{g_{V\gamma
P}(q^2)}{M_V}\epsilon_{\mu\nu\alpha\beta}\partial^\mu
V^\nu\partial^\alpha A^\beta P \ee where $V^\nu(=\rho, \ \omega, \
\phi, \ J/\psi, \ \psi^\prime\dots)$ and $A^\beta$ are the vector
meson and EM field, respectively; $M_V$ is the vector meson mass;
$\epsilon_{\mu\nu\alpha\beta}$ is the anti-symmetric Levi-Civita
tensor. The coupling constant $g_{V\gamma P}(q^2)$ is off-shell
and involves a form factor due to the virtuality of the photon. It
can be expressed as \be g_{V\gamma P}(q^2)=g_{V\gamma P}(0){\cal
F}(q^2) \ , \ee where $g_{V\gamma P}(0)$ is the on-shell coupling
and can be determined by vector meson radiative
decays~\cite{Zhao:2006gw,Li:2007ky}, e.g. $\omega\to\gamma\pi^0$
and $\phi\to \gamma\pi^0$.

\begin{figure}
 \begin{center}
\epsfig{file=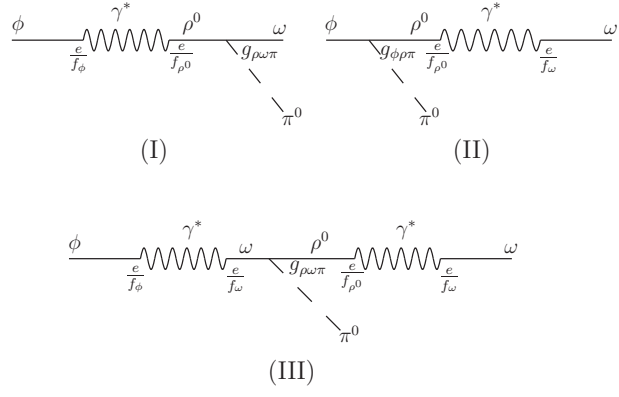, width=14cm,height=20cm}\vspace{-12cm}
 \caption{Schematic diagrams for the EM transitions in $\phi\to \omega\pi^0$.}
 \protect\label{fig-1}
 \end{center}
 \end{figure}

In the VMD model, we can decompose the virtual photon by a sum of
vector mesons as shown by Fig.~\ref{fig-1}. The amplitude for
Process-I (i.e. Fig.~\ref{fig-1}(I)) can be expressed as
\bea\label{em-1}
M_{fi}^{EM-I} & =& \sum_V
\frac{e}{f_V}\frac{M_V^2}{M_\phi^2-M_V^2+iM_V\Gamma_V}\frac{e}{f_\phi}\frac{g_{\omega
V\pi}}{M_\omega}  \varepsilon_{\alpha\beta\mu\nu}
 p_\omega^\alpha\varepsilon_\omega^\beta
 p_\phi^\mu\varepsilon_\phi^\nu \ ,
\eea
where $g_{\omega V\pi}$ is the $VVP$ strong coupling constant, and
$\Gamma_V$ is the total width of the intermediate vector meson.
This gives
\be
g_{V\gamma P}(q^2)=g_{V\gamma P}(0){\cal F}(q^2)=  \sum_V
g_{\omega V\pi}
\frac{e}{f_V}\frac{M_V^2}{M_\phi^2-M_V^2+iM_V\Gamma_V} \ ,
\ee
which relates the on-shell coupling $g_{V\gamma P}(0)$ to an
off-shell coupling with form factors.

Similarly, the transition matrix element for Process-II
(Fig.~\ref{fig-1}(II)) can be written as
\be\label{em-2}
M_{fi}^{EM-II} =\sum_V \frac{e}{f_V}
\frac{M_V^2}{M_\omega^2-M_V^2+iM_V\Gamma_V}\frac{e}{f_\omega}\frac{g_{\phi
V\pi}}{M_\phi}  \varepsilon_{\alpha\beta\mu\nu}
 p_\omega^\alpha\varepsilon_\omega^\beta
 p_\phi^\mu\varepsilon_\phi^\nu \ ,
\ee
where $g_{\phi V\pi}$ is again the strong coupling constant.

In the VMD framework, it also allows contributions from
Process-III (Fig.~\ref{fig-1}(III)) of which the expression is
\be\label{em-3}
M_{fi}^{EM-III}= \sum_{V_1 V_2} \frac{e}{f_{V_1}}\frac{e}{f_{V_2}}
\frac{M_{V_1}^2}{M_\phi^2-M_{V_1}^2+iM_{V_1}\Gamma_{V_1}}
\frac{M_{V_2}^2}{M_\omega^2-M_{V_2}^2+iM_{V_2}\Gamma_{V_2}}
\frac{e}{f_\omega}\frac{e}{f_\phi}\frac{g_{V_1 V_2\pi}}{M_{V_1}}
\varepsilon_{\alpha\beta\mu\nu}
 p_\omega^\alpha\varepsilon_\omega^\beta
 p_\phi^\mu\varepsilon_\phi^\nu
\ ,
\ee
where $V_1$ and $V_2$ are intermediate vector mesons which are
different from $\omega$ and $\phi$ when they are connected to
these two states by the virtual photon. However, since we adopt
experimental data for $\phi\to \rho^0\pi^0$ in Process-II to
determine the $g_{\phi\rho^0\pi^0}$ coupling, contributions from
Process-III will have been included in Process-II. Nonetheless, we
note in advance that exclusive contributions from Process-III are
negligibly small. Therefore, we will only concentrate on the first
two processes in this study.

The following points can be made about $\phi\to\omega\pi^0$:

i) We argue that the dominant contributions are from $\rho^0$ in
this kinematics. Contributions from higher states will be
relatively suppressed because their masses are larger than the
virtuality of the photon. Other suppressions from the $V\gamma^*$
and $VVP$ couplings are also expected. Basically, those higher
vector mesons are farther away from the $\phi$ and $\omega$ masses
than the $\rho^0$. We thus make an approximation of
Eqs.~(\ref{em-1}) and (\ref{em-2}) by considering only the $\rho$
meson contributions:
\bea
M_{fi}^{EM} &=& M_{fi}^{EM-I}+ M_{fi}^{EM-II}
\nonumber\\
&\equiv & \frac{\tilde{g}_{EM}}{M_\phi}
\varepsilon_{\alpha\beta\mu\nu}
 p_\omega^\alpha\varepsilon_\omega^\beta
 p_\phi^\mu\varepsilon_\phi^\nu
\ ,
\eea
where the EM coupling $\tilde{g}_{EM}$ has a form:
\bea
\tilde{g}_{EM} &\simeq & \frac{e}{f_\rho}\left[\frac{e}{f_\phi}
\left(\frac{M_\phi}{M_\omega}\right)\frac{M_\rho^2}{M_\phi^2-M_\rho^2+iM_\rho\Gamma_\rho}
g_{\omega\rho^0\pi^0}+
\frac{e}{f_\omega}\frac{M_\rho^2}{M_\omega^2-M_\rho^2+iM_\rho\Gamma_\rho}
g_{\phi\rho^0\pi^0}\right]  \ ,
\eea
with $\Gamma_\rho$ and $\Gamma_\omega$ the total widths of
$\rho^0$ and $\omega$, respectively.

ii) The vector-meson-photon couplings, $e/f_V$, can be determined
by $V\to e^+ e^-$:
\be
\frac{e}{f_V}=\left[\frac{3\Gamma_{V\to e^+ e^-}}{2\alpha_e |{\bf
p}_e|}\right]^{1/2} ,
\ee
where  $|{\bf p}_e|$ is the electron three-momentum in the vector
meson rest frame, and $\alpha_e=1/137$ is the fine-structure
constant.

iii) The coupling, $g_{\omega\rho^0\pi^0}^2\simeq 85$, can be well
determined by either $\omega\to \gamma\pi^0$ or $\omega\to\pi^0
e^+ e^-$~\cite{pdg2006} in the same framework.

iv) For $g_{\phi\rho^0\pi^0}$, the KLOE measurement suggests that
$\phi\to \rho\pi\to \pi^+\pi^-\pi^0$ has a weight of 0.937 in
$\phi\to \pi^+\pi^-\pi^0$~\cite{Aloisio:2003ur}. This gives
\be
0.937 \times\Gamma^{exp}_{\phi\to \rho\pi+\pi^+\pi^-\pi^0}
=\frac{|{\bf p}|^3}{12\pi
M_\phi^2}(g_{\phi\rho^0\pi^0}+g_{\phi\rho^+\pi^-}+g_{\phi\rho^-\pi^+})^2
 \ ,
\ee
with $|{\bf p}|$ denoting the three-vector momentum of the final
state meson in the $\phi$-rest frame. It is reasonable to assume
$g_{\phi\rho^0\pi^0}=g_{\phi\rho^+\pi^-}=g_{\phi\rho^-\pi^+}$,
Thus, the coupling constant can be determined:
$g_{\phi\rho^0\pi^0}=0.68$.

On the other hand, the coupling $g_{\phi\rho^0\pi^0}$ can be
extracted in $\phi\to \gamma \pi^0$ by assuming that the $\rho^0$
is the dominant contribution to the form factor. This leads to
\be
g_{\phi\rho^0\pi^0}=\left(\frac{ 12\pi M_\phi^2\Gamma_{\phi\to
\gamma\pi^0}}{|{\bf
p}|^3(e/f_\rho)^2}\frac{(M_\rho^2+\Gamma_\rho^2)}{M_\rho^2}\right)^{1/2}\simeq
0.68 \ ,
\ee
where the $\rho$ meson width is included. These two results are in
excellent agreement with each other and highlight the necessity of
considering the width effects of the $\rho^0$ pole in the form
factor. Also, this evidently shows that the $\rho^0$ pole is the
dominant contribution in the $\phi$ meson radiative decays, and
the VMD approach indeed provides a reliable description of the EM
transitions in $\phi\to \omega\pi^0$.

In the above treatment all the couplings are determined by
experimental data and there is no free parameter in the
calculation of the EM decay couplings.

\subsection{Intermediate $K\bar{K}(K^*)+c.c.$ loop}

As discussed in the Introduction that one, in principle, should
include all the possible intermediate meson exchange loops in the
calculation. In reality, the break-down of the local quark-hadron
duality allows us to pick up the leading contributions as a
reasonable approximation~\cite{Lipkin:1986bi,Lipkin:1986av}. In
the $\phi$ meson decay, the leading branching ratio is via
$\phi\to K\bar{K}$, which makes the intermediate $K\bar{K}$
rescattering via $K^*$ exchange a dominant contribution. Apart
from this, $\phi K^*\bar{K}$ coupling is sizeable in the SU(3)
flavor symmetry which also makes the intermediate
$K\bar{K^*}+c.c.$ rescattering via kaon and/or $K^*$ exchange
important contributions in $\phi\to\omega\pi^0$. Contributions
from higher mass states turn to be suppressed at the $\phi$ mass
region. We take this as a reasonable approximation in this work,
and formulate the contributions from i) intermediate
$K\bar{K}(K^*)$ loop; ii) intermediate $K\bar{K^*}(K)$ loop; and
iii) intermediate $K\bar{K^*}(K^*)$ loop.

The transition amplitude for $\phi \to \omega \pi^0$ via an
intermediate meson loop can be expressed as follows:
\bea\label{amp-loop-1}
 M_{fi}=\int \frac {d^4p_2}{(2\pi)^4}\sum_{K^* pol}
 \frac {T_1T_2T_3}{a_1a_2a_3}{\cal F}(p_2^2) \ .
\eea
For  $K\bar{K}(K^*)$, the vertex functions are
\be
\left\{\begin{array}{ccc}
 T_1 &\equiv& ig_1(p_1-p_3)\cdot \varepsilon_\phi \\
 \nonumber
 T_2&\equiv& \frac
 {ig_2}{M_\omega}\varepsilon_{\alpha\beta\mu\nu}p_\omega^\alpha\varepsilon_\omega^\beta
 p_2^\mu\varepsilon_2^\nu \\
 T_3&\equiv& ig_3(p_\pi+p_3)\cdot \varepsilon_2\end{array}\right.
 \ee
where $g_1$, $g_2$, and $g_3$ are the coupling constants at the
meson interaction vertices (see Fig. \ref{fig-1}). The four
vectors, $p_\phi$, $p_\omega$, and $p_{\pi^0}$ are the momenta for
the initial $\phi$ and final state $\omega$ and $\pi$ meson; The
four-vector momentum, $p_1$, $p_2$, and $p_3$ are for the
intermediate mesons, respectively, while $a_1=p_1^2-m_1^2,
a_2=p_2^2-m_2^2$, and $a_3=p_3^2-m_3^2$ are the denominators of
the propagators of intermediate mesons.

The form factor ${\cal F}(p^2)$, which takes care of the off-shell
effects of the exchanged particles, is usually parameterized as
\be
 {\cal F}(p^2) = \left(\frac {\Lambda^2 - m^2}{\Lambda^2 - p^2}\right)^n,
\ee
where $n=0, 1, 2$ correspond to different treatments of the loop
integrals.

The coupling constants for the charged and neutral meson
interactions are denoted by subscription ``c" and ``n",
respectively. In the charged meson exchange loop, coupling
$g_{1c}$ can be determined by the experimental data for $\phi\to
K^+K^- + c.c$,
\bea
 g_{1c}^2=\frac {6\pi M_\phi^2}{|{\bf P}_{1c}|^3}\Gamma_{\phi \to
 K^+K^- +c.c. } \ ,
\eea
where $\Gamma_{\phi \to K^+K^-+c.c} =(49.2\pm 0.6)\%\times
\Gamma_{tot}$~\cite{pdg2006}. For the neutral channel, $g_{1n}$ is
determined by $\phi\to K^0{\bar
 K}^0 + c.c.$ for which we adopt $\Gamma_{\phi \to K_SK_L}=(34.0\pm 0.5)\%\times
\Gamma_{tot}$~\cite{pdg2006} to derive:
\bea
 g_{1n}^2=\frac {6\pi M_\phi^2}{|{\bf P}_{1n}|^3}\Gamma_{\phi \to
 K_SK_L} \ .
\eea

The coupling constant $g_{3c}$  and $g_{3n}$ can be deduced
through the decay $K^*\to K\pi$. For example, $g_{3n}$ is
determined by $K^{* 0}\to K^0\pi^0$:
\bea
 g_{3n}^2=g_{K^{* 0} K^0\pi^0}^2 = \frac{6\pi M_{K^{* 0}}^2}{|{\bf P}|^3}
{\Gamma_{K^{* 0} \to K^0 \pi^0}} \ .
\eea
It shows that within the precision of the experimental data for
$K^{* 0}\to K^0\pi^0$ and $K^{*\pm}\to K^\pm \pi^\mp$, coupling
$g_{K^{* 0} K^0\pi^0}$ has the same value as $g_{K^{*\pm} K^\pm
\pi^\mp}$. The extracted values are listed in Table~\ref{tab-1}.

The relative signs between the couplings are determined by the
SU(3) flavor symmetry relations~\cite{tornqvist}: \be g_{3c} =
-g_{3n}=g_{K^{* -} K^-\pi^0}=- g_{K^{* +} K^+\pi^0} =g_{K^{* 0}
K^0\pi^0} =
 - g_{\bar {K}^{* 0} \bar {K}^0\pi^0}  \ .
\ee
Note that the above equation is to illustrate the relative signs
instead of the values for the coupling constants.

The coupling constant $g_2$ cannot be directly derived from
experiment. But it can be related to the $\omega\rho^0\pi^0$
coupling via the SU(3) flavor symmetry:
\be
g_{2c}= g_{2n}=g_{\omega  K^{* -} K^+}=g_{\omega K^{* +} K^-}
  =g_{\omega {\bar  K}^{* 0} K^0}=g_{\omega  K^{* 0} {\bar K}^0}= g_{\omega\rho^0\pi^0}/2,
\ee where, again, the relative signs between the charged and
neutral couplings are determined by Ref.~\cite{tornqvist}.

With the couplings determined as the above, one can see that a
relative sign arises between the amplitudes for the charged and
neutral meson exchange loops. We then distinguish these two
amplitudes as follows:
\be\label{amp-cn}
M_{fi}\equiv M_{fi}^c+M_{fi}^n \ ,
\ee
where $M_{fi}^c$ and $M_{fi}^n$ have similar structures except
that the couplings and masses involving the intermediate charged
and neutral mesons are different due to the isospin symmetry
violations. The nonvanishing cancellation thus can contribute to
the isospin-violating branching ratios.

To proceed, we treat the loop integral in two different ways.
Firstly, we apply an on-shell approximation (Cutkosky rule) for
the intermediate $K\bar{K}$, which will reduce the loop
integration into an integral over the azimuthal angles defined by
${\bf p}_3$ relative to ${\bf p}_\pi$. This approximation picks up
the imaginary part of the transition amplitude, and with $n=0, \
1, \ 2$, we can examine the effects from the form factors.
Disadvantage of this treatment is that for intermediate mesons of
which the mass threshold is above the $\phi$ mass, their
contributions to the imaginary (absorptive) part vanish though
their contributions to the real (dispersive) part may be sizeable.
Because of this, we also consider the loop integrals including the
dispersive part in a Feynman integration. To kill the ultraviolet
divergences, we include the form factors with $n=1$ and $2$ for a
monopole and dipole, respectively. Below are the details.

\subsubsection{ Integrations with on-shell approximation}

By applying the Cutkosky rule to the loop integration, we can
reduce the transition amplitude (e.g. for the charged meson loop)
to be:
\be
 M_{fi}^c = \frac {|{\bf p}_{3c}|} {32\pi^2M_\phi} \int d\Omega
 \frac {T_c {\cal F}(P_{2c}^2)}{p_{2c}^2 - m_{2c}^2} \ ,
\ee with \be T_c\equiv (T_1T_2T_3)_c=\frac {ig_{1c}g_{2c}g_{3c}}
{M_\omega} 4\varepsilon_{\alpha\beta\mu\nu}
 \varepsilon_\omega^\alpha p_{3c}^\beta p_\pi^\mu p_\omega^\nu \varepsilon_\phi
 \cdot p_{3c} \ .
\ee
The integration is over the azimuthal angles of the momentum ${\bf
p}_{3c}$ relative to the momentum of the final state $\pi$ meson.
The kinematics are defined as $p_\omega = (E_\omega, 0, 0, |{\bf
P}_\omega|)$, $p_\pi = (E_\pi, 0, 0, -|{\bf P}_\omega|)$, and
$p_{2c}^2=(p_{3c} -p_\pi)^2 = M_\pi^2+m_{3c}^2-2E_\pi E_{3c} +
2|{\bf P}_\pi||{\bf p}_{3c}| {\cos \theta}$.

Similarly, we obtain the amplitude for the neutral meson loop:
\be
 M_{fi}^n = \frac {|{\bf p}_{3n}|} {32\pi^2M_\phi} \int d\Omega
 \frac {T_n {\cal F}(p_{2n}^2)}{p_{2n}^2 - m_{2n}^2} \ ,
\ee with
\be T_n\equiv (T_1T_2T_3)_n=\frac {ig_{1n}g_{2n}g_{3n}}
{M_\omega} 4\varepsilon_{\alpha\beta\mu\nu}
 \varepsilon_\omega^\alpha p_{3n}^\beta p_\pi^\mu p_\omega^\nu \varepsilon_\phi
 \cdot p_{3n} \ .
\ee
Note that the momenta and masses for the intermediate states are
different between the charged and neutral cases as denoted by the
subscription ``c" and ``n", respectively.

The nonvanishing amplitudes require the vector meson polarizations
to be taken as either
$(\varepsilon_\omega,\varepsilon_\phi)=(+,-)$ or $(-,+)$. We then
obtain
\be\label{kk-on-shell} M_{fi}(+,-) =  -M_{fi}(-,+) =- \frac
{g_1g_2g_3 |{\bf p}_3|^3 |{\bf P}_\omega|} {8\pi M_\omega} {\cal
I} \ , \ee where \be {\cal I}\equiv\int \frac {\sin^2\theta {\cal
F}(P_2^2)}{p_2^2-m_2^2} \sin\theta d\theta \ .
\ee

(i) With no form factor, i.e., ${\cal F}(p_2^2)=1$, the integral
becomes:
\be
 {\cal I} =
\frac {1} {A_s}\left[\frac {2}{A^2} + \frac {A^2-1}{A^3}\log
{\frac {1+A}{1-A}}\right] \ .
\ee

(ii) With a monopole form factor, i.e., ${\cal
F}(p_2^2)=(\Lambda^2 -m^2_2)/(\Lambda^2 -
 p_2^2)$,  the integral becomes:
\bea
{\cal I} = \frac {m_2^2-\Lambda^2}{A_sB_s} \left[-\frac {2} {AB} +
\frac {A^2-1} {A^2(A-B)}\log{\frac {1+A}{1-A}} +\frac {1-B^2}
{B^2(A-B)}\log{\frac
 {1+B}{1-B}}\right]\ .
\eea

(iii) With a dipole form factor, i.e., ${\cal
F}(p_2^2)=[(\Lambda^2 -m^2_2)/(\Lambda^2 - p_2^2)]^2$, the
integral becomes
\bea
{\cal I} = \frac {(m_2^2-\Lambda^2)^2}{A_sB_s^2(A-B)^2}
\left[-\frac {2B(A-B)(B^2-1)} {B^2(1-B^2)} + \frac {A^2-1}
{A}\log{\frac {1+A}{1-A}} -\frac {AB^2-2B+A} {B^2}\log{\frac
 {1+B}{1-B}}\right] \ .
\eea
The kinematic functions are defined as
\bea
 A_s&=&M_\omega^2+m_1^2-2E_1E_\omega -m_2^2, \\ \nonumber
 B_s&=&M_\omega^2+m_1^2-2E_1E_\omega -\Lambda^2,\\ \nonumber
 A&=&-2|{\bf p}_1||{\bf P}_\omega|/A_s, \\
 B&=&-2|{\bf p}_1||{\bf P}_\omega|/B_s .
\eea

\subsubsection{ Feynman integrations with form factors}

With the form factors, the ultraviolet divergence in the Feynman
integration can be avoided. For the charged meson loop as an
example, the integral has an expression:
\be
{\cal M}_{fi}^c  =  \int \frac {d^4p_{2c}} {(2\pi)^4} \sum _{K^*
pol } \frac  {[ig_{1c} (p_{1c} - p_{3c} ) \cdot
 \varepsilon_\phi] [\frac
 {ig_{2c}}{M_\omega}\varepsilon_{\alpha\beta\mu\nu}p_\omega^\alpha\varepsilon_\omega^\beta
 p_{2c}^\mu \varepsilon_2^\nu] [ig_{3c}(p_\pi + p_{3c})\cdot
 \varepsilon_2]}{(p_{1c}^2-m_{1c}^2)(p_{3c}^2-m_{3c}^2)(p_{2c}^2-m_{3c}^2)}
{\cal F}(p_{2c}^2) \ .
\ee
With a monopole form factor, we have
\be\label{monopole-loop-1}
 {\cal M}_{fi}^c = -\frac {g_{1c} g_{2c} g_{3c}}{M_\omega}
 \varepsilon_{\alpha\beta\mu\nu}
 p_\omega^\alpha\varepsilon_\omega^\beta p_\phi^\mu\varepsilon_\phi^\nu
 \int^{1}_{0}dx\int^{1-x}_{0}dy
\frac{2}{(4\pi)^2}\log \frac {\triangle
(m_{1c},m_{3c},\Lambda)}{\triangle (m_{1c},m_{3c},m_{2c})} \ ,
\ee
while with a dipole form factor, we have
\bea\label{dipole-loop-1}
 {\cal M}_{fi}^c &=& -\frac {g_{1c} g_{2c} g_{3c}}{M_\omega}
 \varepsilon_{\alpha\beta\mu\nu}
 p_\omega^\alpha\varepsilon_\omega^\beta p_\phi^\mu\varepsilon_\phi^\nu
 \int^{1}_{0}dx\int^{1-x}_{0}dy
 \frac{2}{(4\pi)^2}\left[\log \frac {\triangle (m_{1c},m_{3c},\Lambda)}{\triangle (m_{1c},m_{3c},m_{2c})}\right.  \\
 & &\left.- \frac {y(\Lambda^2 -m_{2c}^2)}{\triangle (m_{1c},m_{3c},\Lambda )}\right]
\eea
where the function $\Delta$ is defined as
\bea
 \Delta(a,b,c)&\equiv & M_{\omega}^{2}(1-x-y)^2-(M_\phi^2-M_\omega^2-M_\pi^2)(1-x-y)x+M_{\pi}^2 x^2
 -(M_{\omega}^{2}-a^{2})(1-x-y)\nonumber\\&&
 -(M_{\pi}^{2}-b^{2})x+yc^2 \ .
\eea

Expressions for $M_{fi}^n$ are essentially the same as $M_{fi}^c$
with $g_{1c,2c,3c}$ and $m_{1c,2c,3c}$ replaced by $g_{1n,2n,3n}$
and $m_{1n,2n,3n}$, and we do not repeat them here in order to
save space.

\subsection{Intermediate $K\bar{K^*}(K)+c.c.$ loop}

\begin{figure}
\begin{center}
\epsfig{file=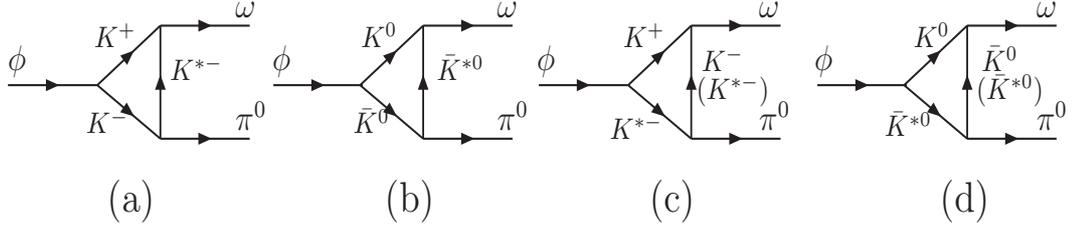, width=14cm,height=3cm}
%\vspace{-16cm}
\caption{Schematic picture for the decay of $\phi \to \omega\pi^0$
via $K\bar{K}(K)$, $K\bar{K^*}(K)$ and $K\bar{K^*}(K^*)$
intermediate meson loops.} \protect\label{fig-2}
\end{center}
\end{figure}

As shown by Fig.~\ref{fig-2}, the vertex functions for the
$K\bar{K^*}(K)+c.c.$ loop are
\be
\left\{ \begin{array}{ccc}
 T_1 &\equiv &\frac {i f_1}{M_\phi}
 \varepsilon_{\alpha\beta\mu\nu}
 p_\phi^\alpha \varepsilon_\phi^\beta p_3^\mu \varepsilon_3^\nu \ , \nonumber \\
 T_2&\equiv& i f_2(p_1-p_2)\cdot \varepsilon_\omega \ , \nonumber\\
 T_3&\equiv & i f_3(p_\pi-p_2)\cdot \varepsilon_3 \ .
 \end{array}\right.
\ee
where $f_{1,2,3}$ are the coupling constants and and ${\cal
F}(p_2^2)$ is the form factor.

Similar to the previous Section, one finds that a relative sign
arises from the charged and neutral meson exchange loops, which
can be distinguished by $M_{fi}\equiv M_{fi}^c+ M_{fi}^n $. Thus,
we have the expression for the charged amplitude with a monopole
form factor:
\be\label{monopole-loop-2}
 {\cal M}_{fi}^c = \frac {f_{1c} f_{2c} f_{3c}}{M_\omega}
 \varepsilon_{\alpha\beta\mu\nu}
 p_\omega^\alpha\varepsilon_\omega^\beta p_\phi^\mu\varepsilon_\phi^\nu
 \int^{1}_{0}dx\int^{1-x}_{0}dy
\frac{2}{(4\pi)^2}\log \frac {\triangle
(m_{1c},m_{3c},\Lambda)}{\triangle (m_{1c},m_{3c},m_{2c})} \ ,
\ee
and with a dipole form factor:
\bea\label{dipole-loop-2}
 {\cal M}_{fi}^c &=& \frac {f_{1c} f_{2c} f_{3c}}{M_\omega}
 \varepsilon_{\alpha\beta\mu\nu}
 p_\omega^\alpha\varepsilon_\omega^\beta p_\phi^\mu\varepsilon_\phi^\nu
 \int^{1}_{0}dx\int^{1-x}_{0}dy
 \frac{2}{(4\pi)^2}\left[\log \frac {\triangle (m_{1c},m_{3c},\Lambda)}{\triangle (m_{1c},m_{3c},m_{2c})}\right.  \\
 & &\left.- \frac {y(\Lambda^2 -m_{2c}^2)}{\triangle (m_{1c},m_{3c},\Lambda
 )}\right] \ .
\eea
In the above two equations the intermediate meson masses
$m_{1,2,3}$ are from the $K\bar{K^*}(K)$ loops, which are
different from those in Eqs.~(\ref{monopole-loop-1}) and
(\ref{dipole-loop-1}).

In the $K\bar{K^*}(K)$ loop, the coupling constant $g_{\phi K^*
K}$ is related to $g_{\omega\rho^0\pi^0}$ in the SU(3) flavor
symmetry:
\be
f_{1c}=f_{1n}=g_{\phi K^{*+}K^-}=g_{\phi K^{*-}K^+}=g_{\phi
K^{*0}\bar{K^0}}=g_{\phi
\bar{K^{*0}}K^0}=g_{\omega\rho^0\pi^0}/\sqrt{2} \ ,
\ee
where we neglect the possible differences caused by the isospin
violation between the charged and neutral channel. The reason is
because this loop contributions are negligibly small and such a
differences cannot produce measurable effects. At the $\omega
K\bar{K}$ vertex, the coupling $g_{\omega K\bar{K}}$ can be
related to $\phi K\bar{K}$ by the following relation:
\bea
f_{2c} &=& g_{\omega K^+ K^-}  = -g_{\omega K^-K^+}=g_{\phi
K^+K^-}/\sqrt{2} \ , \nonumber\\
f_{2n} &=& g_{\omega K^0\bar{K^0}} = -g_{\omega \bar{K^0}K^0} =
g_{\phi K^0\bar{K^0}}/\sqrt{2} \ ,
\eea
where we assume that the isospin breaking in the $\omega K\bar{K}$
couplings is similar to that in the $\phi K\bar{K}$ ones.

 The absolute values of the coupling constants are
listed in Table~\ref{tab-1}.

\subsection{Intermediate $K\bar{K^*}(K^*)+c.c.$ loop}

We also consider the transition amplitude from the intermediate
$K\bar{K^*}(K^*)+c.c.$ loop (Fig.~\ref{fig-2}), which can be
expressed the same form as Eq.~(\ref{amp-loop-1}) except that the
vertex functions change to
\be\label{vertex-1b}
\left\{\begin{array}{ccl}
 T_1 &\equiv &\frac {i h_1}{M_\phi}
 \varepsilon_{\alpha\beta\mu\nu}
 P_\phi^\alpha \varepsilon_\phi^\beta p_3^\mu \varepsilon_3^\nu \ , \\
 \nonumber
 T_2&\equiv& \frac {i h_2}{m_2}
 \varepsilon_{\alpha^\prime\beta^\prime\mu^\prime\nu^\prime}
 p_2^{\alpha^\prime} \varepsilon_2^{\beta^\prime} P_\omega^{\mu^\prime} \varepsilon_\omega^{\nu^\prime} \ , \\
 T_3&\equiv & \frac {i h_3}{m_3}
 \varepsilon_{\alpha^{\prime\prime}\beta^{\prime\prime}\mu^{\prime\prime}\nu^{\prime\prime}}
 p_2^{\alpha^{\prime\prime}} \varepsilon_2^{\beta^{\prime\prime}} p_3^{\mu^{\prime\prime}} \varepsilon_3^{\nu^{\prime\prime}}
\end{array}\right.
 \ee
where $f_{1,2,3}$ are the coupling constants and ${\cal F}(p_2^2)$
is the form factor.

Similar to the above Sections, there is a relative sign arise from
the charged and neutral meson exchange loops, i.e. $ M_{fi} \equiv
M_{fi}^c + M_{fi}^n $, and we only give here the expressions for
the charged amplitude with a monopole and dipole form factor
respectively,
\be
M_{fi}^c = \frac {h_{1c}h_{2c}h_{3c} } {M_\phi
m_{2c}m_{3c}}\epsilon_{\alpha\beta\mu\nu} p_\omega^\alpha
\epsilon_\omega^\beta p_\phi^\mu  \epsilon_\phi^\nu \int_0^1dx
\int_0^{1-x}dy \int_0^{1-x-y} dz   \frac {2} {(4\pi)^2} [\frac {A}
{\triangle_1} - \frac {B} { 2\triangle_1^2}] \ ,
\ee
and
\be
M_{fi}^c = -\frac {h_{1c}h_{2c}h_{3c} } {M_\phi
m_{2c}m_{3c}}\epsilon_{\alpha\beta\mu\nu} p_\omega^\alpha
\epsilon_\omega^\beta p_\phi^\mu  \epsilon_\phi^\nu \int_0^1dx
\int_0^{1-x}dy \int_0^{1-x-y} dz \frac {2} {(4\pi)^2} [\frac {A}
{\triangle_1^2} - \frac {B} { \triangle_1^3}] \ ,
\ee
with
\bea
A &= &\frac {1} {4} (2x+\frac {3} {2}z -1) (M_\phi^2-M_\omega^2 -
M_{\pi^0}^2) +\frac {1}{2}xM_{\pi^0}^2 + \frac {1} {4}z M_\omega^2
\ ,\nonumber\\
B &= & (x+z-1)xz[M_\omega^2M_{\pi^0}^2 -\frac {1}
{4}
(M_\phi^2-M_\omega^2 - M_{\pi^0}^2)^2] \ , \nonumber\\
\triangle_1 &= & x^2 M_{\pi^0}^2 +z^2M_\omega^2 -x z
(M_\phi^2-M_\omega^2 - M_{\pi^0}^2)-z(M_\omega^2 - M_{1c}^2)
\nonumber\\
&& +yM_{2c}^2 - x(M_{\pi^0}^2-M_{3c}^2)+ (1-x-y-z)\Lambda^2 \ .
\eea

In this transition loop the intermediate meson masses $m_{1,2,3}$
correspond to $K$, $\bar{K^*}$ and $k^*$. Quantities $h_{1,2,3}$
denote the corresponding vertex coupling constants with the
relative signs given by:
\be
h_{3c}=-h_{3n}=-g_{{\bar K}^{\ast 0}{{\bar K}^{\ast
0}}{\pi^0}}=-g_{K^{\ast 0}{K^{\ast 0}}{\pi^0}}=-g_{K^{\ast
+}{K^{\ast +}}{\pi^0}}=-g_{K^{\ast -}{K^{\ast
-}}{\pi^0}}=g_{\omega\rho^0\pi^0}/2 \ .
\ee

\section{Numerical Results}

\subsection{Branching ratios from EM decay transition}

The $\phi$ meson EM decay turns to be very sensitive to the
$\rho^0$ mass pole and decay width in the VMD model. This is
because their masses are close to each other. As a test, in the
infinitely-narrow-width limit, i.e. $\Gamma_\rho=\Gamma_\omega=0$
GeV, the branching ratio turns out to be overestimated:
$BR^{EM}=1.46\times 10^{-4}$, which is more than two times of the
experimental value. This may not be surprising since one should
adopt the mass eigenstates in the calculation instead of the
isospin eigenstates in degenerate perturbation theory. Therefore,
we apply the experimental data for the intermediate vector meson
masses and widths in the calculation.

With the width of the $\rho$ meson included, we obtain
$BR^{EM}=1.68\times 10^{-5}$, with $M_\rho=775.9$ MeV and
$\Gamma_\rho=143.9$ MeV~\cite{Aloisio:2003ur}. With the PDG
average, i.e. $M_\rho=775.5$ MeV and $\Gamma_\rho=149.4$ MeV, we
have $BR^{EM}=1.67\times 10^{-5}$. This explicitly shows an
important role played by the $\rho$ meson.

We also examine the relative strength between Process-I and II.
Their exclusive contributions to the branching ratios are
$BR^{EM-I}=1.45\times 10^{-5}$ and $BR^{EM-II}=4.56\times
10^{-7}$, respectively, which shows that Process-I is dominant
over II in the $\phi$ decay.

The above results suggest that the EM transition alone cannot
account for the observed branching ratio for $\phi\to
\omega\pi^0$. We hence need to look at the contributions from the
intermediate meson exchanges.

\subsection{Branching ratios from hadronic loop under on-shell approximation}

Under the on-shell approximation only the intermediate $K\bar{K}$
will contribute since the threshold of any other strange meson
pairs will be above the $\phi$ mass.

Without the form factor, the branching ratio from $K\bar{K}(K^*)$
loop is $3.02\times 10^{-6}$. This number is much smaller than the
EM contributions. Apart from the significant cancellations between
the charged and neutral channel amplitudes, another reason is
because of the kinematic suppression on the absorptive amplitudes,
i.e. the intermediate $K\bar{K}$ is close to the $\phi$ mass.
Similar phenomena are observed in $J/\psi\to \gamma f_0(1810)\to
\gamma \omega\phi$ at the higher mass tail of the
$f_0(1810)$~\cite{zhao-zou-f0(1810)}. At least it is reasonable to
understand that contributions from near-threshold intermediate
meson rescattering are limited in the on-shell approximation.

\begin{figure}
\begin{center}
\epsfig{file=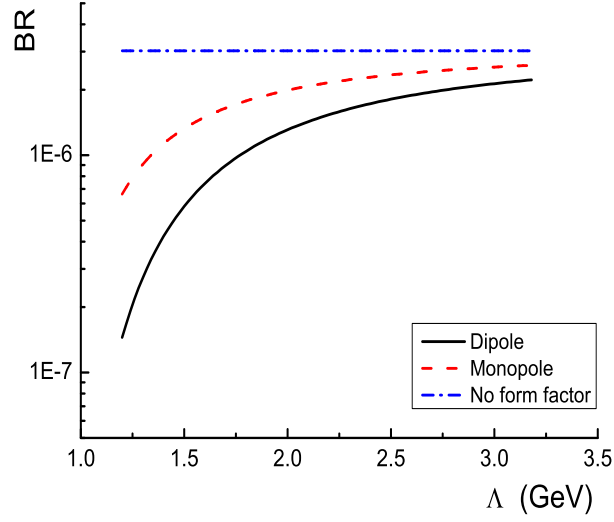, width=9cm,height=8cm} \caption{ The
$\Lambda$-dependence of the $K\bar{K}(K^*)$ loop contributions in
the on-shell approximation. The dot-dashed, dashed and solid curve
denote different considerations for the form factors, i.e. no form
factor, monopole and dipole, respectively.} \protect\label{fig-3}
\end{center}
\end{figure}

In order to investigate the role played by the form factors, we
present the calculation results in Fig.~\ref{fig-3} for three
cases: i) The hadronic loop has a dipole form factor (solid
curve); ii) The hadronic loop has a monopole form factor (dashed
curve); and iii) no form factors are included (dot-dashed line).
It is easy to understand that under the on-shell approximation the
calculation without the form factors for the hadronic loops will
have the largest contributions to the branching ratio. In
contrast, the inclusion of a monopole form factor suppresses the
hadronic loop contributions, and a dipole form factor leads to the
most suppressions. These three results then converge to the same
value when $\Lambda\to \infty$ as shown in Fig.~\ref{fig-3}.

\begin{figure}
\begin{center}
\epsfig{file=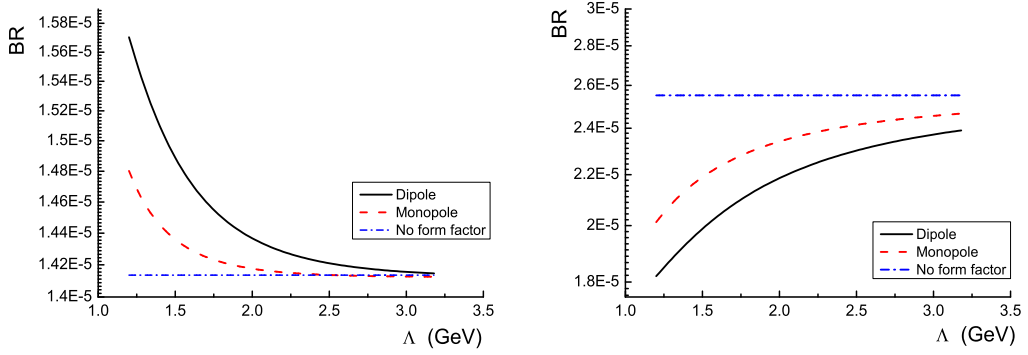, width=16cm,height=7cm} \caption{ The
$\Lambda$-dependence of the sum of the EM and $K\bar{K}(K^*)$ loop
amplitudes in the on-shell approximation. The left panel indicates
results for a destructive addition and the right panel for a
constructive addition. The solid, dashed and dot-dashed curves
denote different considerations for the form factors, i.e. dipole,
monopole and no form factor, respectively. } \protect\label{fig-4}
\end{center}
\end{figure}

The overall results in terms of $\Lambda$ including the EM and
hadronic loop amplitudes are presented in Fig.~\ref{fig-4} for two
different phases, i.e. on the left panel the EM amplitude is out
of phase to the hadronic loop (destructive addition), while on the
right panel these two amplitudes are in phase (constructive
addition). On the left panel the horizontal line reflect the
largest cancellation between the EM and hadronic loop amplitudes
with no form factor suppressions. At small $\Lambda$ region, the
cancellations are small for both monopole and dipole calculations
since the hadronic loop amplitudes are small in both cases as
shown by Fig.~\ref{fig-3}. These three curves smoothly approach
the same value at high $\Lambda$ where the hadronic loop
contributions become negligibly small.

On the right panel the EM amplitude is in phase to the hadronic
loop. In the case that no form factor introduced in the hadronic
loop, the constructive addition of the EM and hadronic loop
amplitudes gives $BR=2.55\times 10^{-5}$. For the monopole and
dipole form factor, the constructive effects increase with
parameter $\Lambda$ since the exclusive hadronic loop
contributions are small in small $\Lambda$ region. It shows by the
dashed and solid curve that the inclusive branching ratios
converge to the dot-dashed curve at large $\Lambda$. In this
constructive addition, the maximum branching ratio is still
smaller than the experimental data, which is a sign for the
underestimate of the hadronic loop contributions in the on-shell
approximation, and implies the need for contributions from the
dispersive part, i.e. from intermediate mesons above the $\phi$
mass.

\subsection{Branching ratios from Feynman integrations}

Note that we are interested in a small effect arising from
cancellations between two sizeable amplitudes. Since the charged
and neutral amplitudes distinguish themselves by the mass
differences between the charged and neutral particles involved in
the loop transition, it makes the behavior of the cancellations
very sensitive to the choice of the cut-off energies. Again, it is
necessary to investigate the $\Lambda$ dependence of the hadronic
loop integrals. We first study the exclusive behaviors of the
$K\bar{K}(K^*)$, $K\bar{K^*}(K)$ and $K\bar{K^*}(K^*)$ loops and
then combine them with the EM transitions to study their
interferences.

\begin{figure}
\begin{center}
\epsfig{file=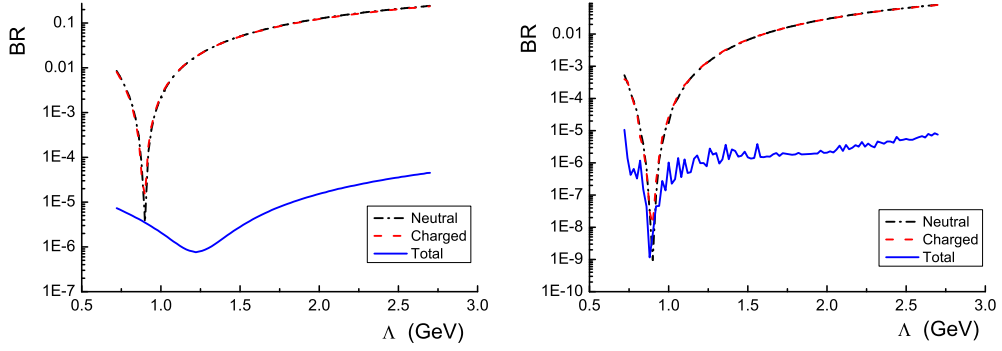, width=16cm,height=7cm} \caption{ The
$\Lambda$-dependence of the $K\bar{K}(K^*)$ loop contributions to
the branching ratio in the Feynman integration. The left panel
indicates results with a monopole form factor, and the right one
with a dipole form factor. The dashed and dot-dashed curves are
contributions from only charged and neutral meson loop,
respectively, while the solid curves are the results after
cancellations between the charged and neutral amplitudes. We note
that the dashed and dot-dashed curves are close to each other and
difficult to distinguish them by sight. } \protect\label{fig-5}
\end{center}
\end{figure}

In Fig.~\ref{fig-5}, the $K\bar{K}(K^*)$ loop in terms of the
cut-off energy $\Lambda$ is illustrated. The left panel is for a
monopole form factor, while the right one is for a dipole type.
The dashed and dot-dashed curves are contributions from the
charged and neutral meson loop, respectively, and the solid curves
are their differences. In fact, the differences between the dashed
and dot-dashed curves are so small that it is hard to distinguish
them as shown by the figures. Their cancellations leave only a
small residue quantity accounting for the isospin violation
effects.

The dependence of the details of the cancellations to the cut-off
energy turns out to be more dramatic with a dipole form factor as
shown by the right panel of Fig.~\ref{fig-5}. Although the
integral for both the charged and neutral meson loops has a
well-defined behavior, details of the cancellations as shown by
the solid curve has an oscillatory behavior at small $\Lambda$.
This is understandable since the difference between the charged
and neutral meson loop integrals has a complicated dependence on
the couplings, and the mass differences between the charged and
neutral kaon and $K^*$ in the propagators. For large $\Lambda$,
the integral difference smooths out since $\Lambda$ becomes the
major energy scale.

In Fig.\ref{fig-5} there are dips appearing at small $\Lambda$ for
both monopole and dipole form factors. This is due to the factor
$\Lambda^2-m_{K^*}^2$ in the numerators of the form factors and
the largest cancellation between the charged and neutral meson
loops.

For the $P$-wave $\phi\to \omega\pi^0$ decay, the form factor
favors a dipole behavior with relatively large $\Lambda$ in order
to account for the off-shell effects. Guided by the solid curve on
the right panel of Fig.~\ref{fig-5}, we argue that $\Lambda\simeq
1.5\sim 2$ GeV is appropriate for the hadronic loop contributions.
Also, in this region, the integral difference has a well-defined
smooth behavior. In the case of monopole form factor, to describe
the experimental data, $\Lambda $ must have a relatively smaller
value, i.e. $< 2$ GeV. Otherwise, the branching ratio will be
overestimated. Due to this ambiguity, we leave the value of
$\Lambda$ to be determined by the experimental data.

\begin{figure}
\begin{center}
\epsfig{file=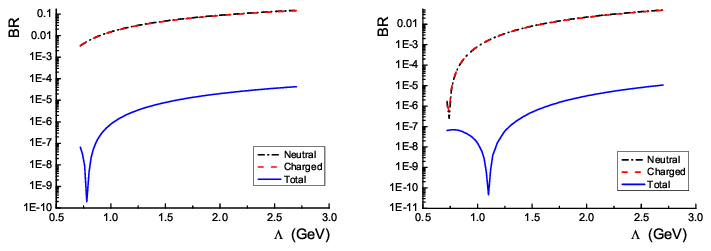, width=16cm,height=7cm} \caption{ The
$\Lambda$-dependence of the $K\bar{K^*}(K)$ loop contributions to
the branching ratio in the Feynman integration. The notations are
similar to Fig.~\ref{fig-5}. Again, we note that the dashed and
dot-dashed curves are difficult to distinguish by sight.}
\protect\label{fig-6}
\end{center}
\end{figure}

The $K\bar{K^*}(K)$ loop contributions are presented by
Fig.~\ref{fig-6} for the monopole and dipole form factors. Similar
to Fig.~\ref{fig-5}, the intermediate charged and neutral meson
loop contributions to the branching ratios are compared with each
other as denoted by the dashed and dot-dashed curves, while the
solid curves are given by their amplitude differences.
Interestingly, the $K\bar{K^*}(K)$ loop contributions turn out to
exhibit a smooth behavior with both monopole and dipole form
factors, and their magnitudes are comparable with the
$K\bar{K}(K^*)$ loop. Again, the dips are related to the factor
$\Lambda^2-m_K^2$ in the numerator of the form factors and the
largest cancellation between the charged and neutral meson loops.

\begin{figure}
\begin{center}
\epsfig{file=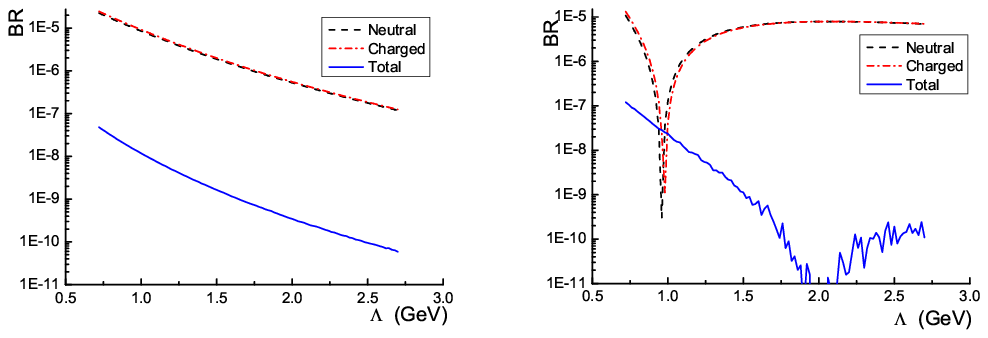, width=16cm,height=7cm} \caption{ The
$\Lambda$-dependence of the $K\bar{K^*}(K^*)$ loop contributions
to the branching ratio in the Feynman integration. The notations
are similar to Fig.~\ref{fig-5}. Again, we note that the dashed
and dot-dashed curves are difficult to distinguish by sight.}
\protect\label{fig-7}
\end{center}
\end{figure}

In  Fig.~\ref{fig-7}, the $\Lambda$-dependence of the exclusive
contributions from the $K\bar{K^*}(K^*)$ loop are presented.
Compared with the other two loops, the exclusive branching ratio
decreases in terms of the increasing $\Lambda$. As a result, its
interferences with other channels around $\Lambda=1.5\sim 2.0$ GeV
turn to be small.

\begin{figure}
\begin{center}
\epsfig{file=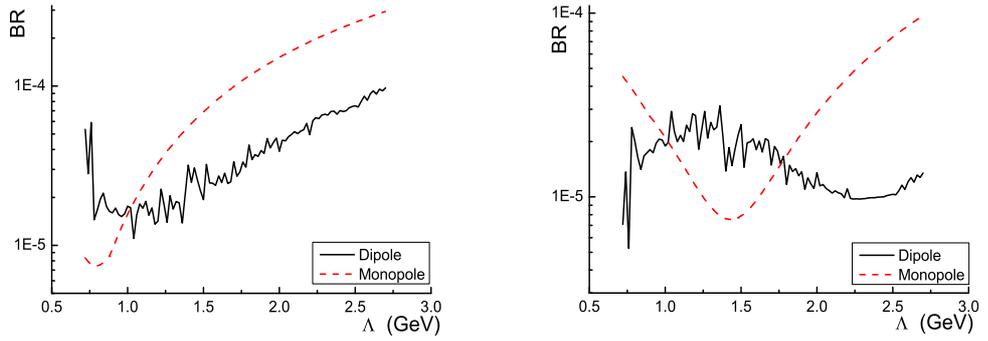, width=16cm,height=7cm} \caption{ The
$\Lambda$-dependence of the constructive (left panel) and
destructive additions (right panel) between the EM and hadronic
loops. The dashed curves denote the results for adopting a
monopole form factor for the hadronic loops, while the solid
curves for adopting a dipole form factor. } \protect\label{fig-8}
\end{center}
\end{figure}

Adding the hadronic loops to the EM amplitude coherently, we
examine two phases in Fig.~\ref{fig-8} in terms of the $\Lambda$,
i.e. constructive (left panel) and destructive additions (right
panel). It shows that with $\Lambda =1.8\sim 2.3$ GeV, the
constructive addition with the dipole form factor for the hadronic
loops gives the branching ratio in agreement with the experimental
data, while with the monopole form factor, $\Lambda$ requires a
range of $1.2\sim 1.5$ GeV. These cut-off energy ranges are
consistent with the commonly accepted values. For a destructive
addition between the EM and hadronic loop amplitudes as shown on
the right panel, we find that the dipole form factor cannot
reproduce the data within $\Lambda=1\sim 2.6$ GeV due to the
significant cancellations between the EM and hadronic loop
transitions. In contrast, with a monopole form factor for the
hadronic loops the destructive addition can still reproduce the
data around $\Lambda=2.3$ GeV. However, this value of $\Lambda$
turns to be out of the commonly accepted range for a monopole
cut-off energy. In this sense, it shows that the data favor a
constructive phase between the EM and hadronic loop amplitudes.

The dipole form factor might be even more preferable. As we have
discussed earlier that the $P$-wave decay will generally favor a
dipole form factor, we hence argue that the constructive addition
between the EM and hadronic loop amplitudes with a dipole form
factor is a favorable mechanism accounting for the experimental
observation of $BR(\phi\to \omega\pi^0)=(5.2_{-1.1}^{+1.3})\times
10^{-5}$~\cite{pdg2006}. In Table~\ref{tab-2}, branching ratios of
the exclusive and coherent (constructively) additions of the EM
and hadronic loops with the dipole and monopole form factors are
listed in comparison with the data.

In comparison with the results given by the on-shell
approximation, it shows that the dispersive part of the loop
transitions plays an important role in reproducing the data.

\subsection{Hadronic loop contributions to the isospin violations in $J/\psi\to
\omega\pi^0$}

Similar to $\phi\to\omega\pi^0$, the decays of $J/\psi\to
\omega\pi^0$ and $\psi^\prime\to\omega\pi^0$ are also isospin
violating processes via DOZI transitions. Their branching ratios
are measured in experiment, i.e. $BR(J/\psi\to\omega\pi^0)=(4.5\pm
0.5)\times 10^{-4}$ and $BR(\psi^\prime\to \omega\pi^0)=(2.1\pm
0.6)\times 10^{-5}$~\cite{pdg2006}, which are not significantly
suppressed compared with $J/\psi(\psi^\prime)\to \phi\eta$,
$\omega\eta^\prime$, etc. An explanation based on vector meson
dominance is provided in Refs.~\cite{Zhao:2006gw,Li:2007ky} where
the branching ratios are fitted by EM transitions with an
appropriate form factor. It also shows that Process-I is the
dominant contributions to the branching ratio while Process-II is
negligibly small.  In this study, a natural question is about the
role played by the hadronic loops and their contributions to the
branching ratios.

Interestingly, $J/\psi\to K^*\bar{K}$ is one of the largest decay
modes, from which relatively large couplings for the $J/\psi
K^*\bar{K}$ vertex can be derived. However, due to the heavy mass
of $J/\psi$, suppressions on the loop amplitudes become crucial.
With the cancellation between the charged and neutral
$K\bar{K}(K^*)$ loops, the hadronic loop contributions to the
branching ratio turn out to be orders of magnitude smaller than
the data. In $\psi^\prime$ decay, the cancellation between the
charged and neutral $K\bar{K^*}(K)$ loops is not as significant as
that in $J/\psi$ where the branching ratios, $BR(J/\psi\to
K^{*+}K^-+c.c.)=(5.0\pm 0.4)\times 10^{-3}$ and $BR(J/\psi\to
K^{*0}\bar{K^0}+c.c.)=(4.2\pm 0.4)\times 10^{-3}$ are close to
each other. In contrast, $BR(\psi^\prime\to
K^{*+}K^-+c.c.)=(1.7^{+0.8}_{-0.7})\times 10^{-5}$ and
$BR(\psi^\prime\to K^{*0}\bar{K^0}+c.c.)=(1.09\pm 0.20)\times
10^{-4}$ have large differences, and have contained significant
contributions from the EM
transitions~\cite{Zhao:2006gw,Li:2007ky}. This favors to maximize
the isospin violation effects in the hadronic loops. However, due
to the suppression from the off-shell form factors, the hadronic
loop contributions will still be negligibly small compared with
the EM transitions.

The numerical calculations show that the branching ratios from the
intermediate $K\bar{K}(K^*)$, $K\bar{K^*}(K)$ and
$K\bar{K^*}(K^*)$ loops in $J/\psi(\psi^\prime)\to \omega\pi^0$
are orders of magnitude smaller than the data. This result
suggests that the EM transition is likely the dominant
isospin-violating process in the vector charmonium decays into
light vector and pseudoscalar mesons. Thus, it enhances the
argument~\cite{Zhao:2006gw,Li:2007ky} that the long-standing
``$\rho\pi$ puzzle" in $J/\psi(\psi^\prime)\to VP$ is mainly due
to the strong destructive interferences from the EM transitions in
$\psi^\prime \to \rho\pi$ which leads to the abnormally small
branching ratio fraction of $BR(\psi^\prime\to
\rho\pi)/BR(J/\psi\to\rho\pi)\simeq 0.2\%$~\cite{pdg2006}.

\section{Summary}

We investigate the isospin-violating mechanisms in $\phi\to
\omega\pi^0$ and $J/\psi\to \omega\pi^0$ by quantifying the EM and
strong transitions as different sources of the isospin violations.
The EM contribution is constrained in the VMD model, and the
hadronic loop contributions is studied by relating them to the
OZI-rule-violating processes. At hadronic level, the OZI-rule
violations are recognized through the nonvanishing cancellations
between the charged and neutral intermediate meson exchange loops.
In another word, the observation of the isospin-violating
branching ratios can be viewed as a consequence of coherent
contributions from the EM transitions and the nonvanishing
cancellations among those intermediate meson exchanges due to the
mass differences between the charged and neutral intermediate
mesons and different couplings to the initial and final state
mesons.

By extracting the vertex coupling information from independent
processes, we can constrain the model parameters and make a
quantitative assessment of the strong isospin violations via
leading $K\bar{K}(K^*)$, $K\bar{K^*}(K)$ and $K\bar{K^*}(K^*)$
loops. It shows that the dispersive part of the hadronic loop
amplitudes have important contributions to the isospin violation
and they produce crucial interferences with the EM transitions
though their exclusive contributions are relatively smaller than
the EM ones in $\phi\to \omega\pi^0$ decay.

We also study the hadronic loop contributions to the isospin
violating decay of $J/\psi(\psi^\prime)\to \omega\pi^0$, and find
that they are negligibly small. This is consistent with our
previous study of the EM transitions in $J/\psi(\psi^\prime)\to
VP$, where we argued that the isospin-violating channels, such as
$\omega\pi^0$, $\rho\eta$, $\rho\eta^\prime$ and $\phi\pi^0$, were
dominated by the EM transitions~\cite{Zhao:2006gw,Li:2007ky}.
However, a caution should be given that in $J/\psi(\psi^\prime)\to
VP$ the $s$-dependence of the intermediate vector meson widths
turns to be a sensitive factor in account of contributions from
light intermediate vector mesons. A coherent study of $e^+ e^-\to
\omega\pi^0$ over a broad range of $s$ is thus strongly desired.

\section*{Acknowledgement}

G. Li would like to thank Y.L. Shen and W. Wang for useful
discussions.  This work is supported, in part, by the U.K. EPSRC
(Grant No. GR/S99433/01), National Natural Science Foundation of
China (Grant No.10675131 and 10521003), and Chinese Academy of
Sciences (KJCX3-SYW-N2).

%%%%%%%%%%%%%%%%%%%%%%%%%%%%%%%%%%%%%%%%%%%%%%%%%%%%%%%
\begin{table}[ht]
\begin{tabular}{|c|c|c|c|c|}
\hline  Coupling constants & $|g_{\phi K\bar{K}}|$ & $|g_{\omega
K^*\bar{K}}|$ & $|g_{K^*K\pi}|(|f_{K^*K\pi}|)$ & $|f_{\phi
K^*\bar{K}}|$  \\ [1ex] \hline Charged kaon coupling &4.49 &4.58
&3.96 &6.48  \\ [1ex] \hline
Neutral kaon coupling &4.62 &4.58 &3.96 &6.48   \\
[1ex] \hline
\end{tabular}
\caption{ The absolute values of coupling constants for the vertex
interactions. Their relative phases are determined by the SU(3)
flavor symmetry.  } \label{tab-1}
\end{table}

\begin{table}[ht]
\begin{tabular}{|c|c||c|c|c|c|c|c|}
\hline & $\Lambda$ (GeV) & EM transition &  $K\bar{K}(K^*)$ &
$K\bar{K^*}(K)$ & $K\bar{K^*}(K^*)$ & Total & Exp. \\ [1ex] \hline
Dipole & 2.14 & 1.66 & 0.23 & 0.33 & $\sim 0.0$ & $5.2\pm 0.2$ & $(5.2^{+1.3}_{-1.1})$ \\
[1ex] \hline
Monopole & 1.38 & 1.66 & 0.14 & 0.56 & $\sim 0.0$ & $5.3\pm 0.5$  & $(5.2^{+1.3}_{-1.1})$ \\
[1ex] \hline
\end{tabular}
\caption{ The exclusive and coherent (constructive) contributions
of the EM and hadronic loops to the $\phi\to \omega\pi^0$
branching ratios with a dipole and monopole form factor. The
experimental data is the world average given by
PDG2006~\cite{pdg2006}. The branching ratios in columns 3-8 have a
unit of $10^{-5}$. The errors estimated in column 7 are due to the
precisions taken for the exclusive branching ratios. }
\label{tab-2}
\end{table}

%%%%%%%%%%%%%%%%%%%%%%%%%%%%%%%%%%%%%%%%

%%%%%%%%%%%%%%

\end{document}